\documentclass[prx,twocolumn,amsmath,amssymb]{revtex4}

\usepackage{color}
\usepackage{bm}
\usepackage{dcolumn}
\usepackage[dvips]{graphicx}
\usepackage{slashed}

\usepackage{ulem}

\newcommand{\1}{\mbox{1}\hspace{-0.25em}\mbox{l}}

\newcommand{\tr}{{\rm tr}\,}

\begin{document}

\preprint{preprint}

\title{
A Dirac fermion model associated with second order topological insulator
}

\author{Takahiro Fukui}
\affiliation{Department of Physics, Ibaraki University, Mito 310-8512, Japan}

\date{\today}

\begin{abstract}
We study topological aspects of a Dirac fermion coupled with a Higgs field associated with 
the lattice model introduced by Benalcazar {\it et al.} which has the topological quadrupole phase. 
Using the index theorem, we show that the index of the Hamiltonian is just given by the winding number of the Higgs field,
implying that a corner state of the lattice model belongs to the same class of the Jackiw-Rossi states 
localized at a vortex.
We also calculate the current density of the Dirac fermion with a symmetry breaking term dependent on time, which is
associated with the dipole pump proposed by Benalcazar {\it et al.}.
We argue that it is indeed a topological current, and the total pumped charge is given by an integer related with the index.
\end{abstract}

\pacs{
}

\maketitle

\section{Introduction}

The bulk-edge correspondence is one of key properties which characterizes topological phases of matter.
While it has been established for the quantum Hall system  \cite{Hatsugai:1993fk}, 
the discovery of the quantum spin Hall effect 
and more generic topological insulators \cite{Kane:2005ab,Kane:2005aa,Qi:2008aa,Hasan:2010fk,Qi:2011kx}
has revealed that the bulk-edge correspondence is valid for wider classes of topological phases.
Weyl semimetals also show unique edge (surface) states called Fermi-arc \cite{Murakami:2007aa,Burkov:2011fk,Wan:2011aa},
which reflect the topological property of the bulk system such as section Chern numbers. 

Recently,  further development has been achieved by Benalcazar, {\it et al.} \cite{Benalcazar:2017ab,Benalcazar:2017aa}.
They have proposed higher order topological insulators which are characterized by $d-D$ dimensional edge (surface) states for 
$d$ dimensional bulk systems. Conventional topological insulators correspond to $D=1$, but 
nontrivial systems with $D>1$ have been successively found and studied extensively
\cite{Liu:2017aa,1611.09680,Langbehn:2017aa,Song:2017aa,Hashimoto:2017aa,Ezawa:2018aa,1801.00437,Schindler:2018aa,1708.03647,Khalaf:2018cr,1804.01531,1804.02794,1805.02598,1805.02831}.
In particular, Khalaf \cite{Khalaf:2018cr}
has pointed out that the corner can be regarded as a topological defect and has
given the classification table of the higher-order topological insulators and superconductors protected by inversion symmemtry, and Trifunovic and Brouwer \cite{1805.02598}
have extended the table
considering generic order-two crystalline (anti-)symmetries.
The corner states in $d=D=2$ dimensional system have indeed observed experimentally in metamaterial circuit systems
\cite{1708.03647}.

In this paper, we study a Dirac fermion model associated with the lattice Benalcazar-Bernevig-Hughes (BBH) model
\cite{Benalcazar:2017ab,Benalcazar:2017aa}.
As the two-dimensional BBH model shows zero-dimensional corner states, it is a second order 
topological insulator with $D=2$.
Although the lattice BBH model includes four Dirac fermions, we pick up one of them and examine the topological 
properties of the single Dirac fermion.

In the next section, we will take the  continuum limit of the BBH model.
Regarding the BBH model as a generalized Wilson-Dirac model \cite{Misumi:2013aa}, we point out that we need not only 
the Dirac fermion at  $\bm k=(0,0)$, but also its doublers
at $\bm k=(\pi,0),(0,\pi),(\pi,\pi)$.
These fermion models have the same structure:
a Dirac fermion model coupled with an O(2) Higgs field, belonging to the symmetry class BDI 
\cite{Altland:1997aa,Schnyder:2008aa}, so we examine the topological property of such a single Dirac fermion.
We show that this model indeed needs corner-like boundaries  to have zero energy states.
In Sec. \ref{s:index}, we discuss the topological property of such corner states using the index theorem.
To this end, we introduce smooth boundaries as a smoothly-varying Higgs field as a function of  the coordinates.
It then turns out that since a corner can be regarded as a point defect, 
a corner state belongs to the same class of the Jackiw-Rossi states  localized at a vortex \cite{JackiwRossi:1981}.
Therefore, the present model belongs to class BDI with a point defect 
in the classification table given 
by Teo and Kane \cite{Teo:2010fk}.
In Sec. \ref{s:pump}, introducing a symmetry breaking term dependent on time, 
we consider a pump of the present Dirac model which corresponds to 
the dipole pump proposed by BBH \cite{Benalcazar:2017ab,Benalcazar:2017aa}.
We calculate the current density for a model with a single defect (a smooth corner as in Sec. \ref{s:index}).
It is shown that the current is indeed topological, and the total pumped charge  is given by an integer 
associated with the index derived in Sec. \ref{s:index}.
When we argue the topological phases of the lattice model, we need to take all Dirac fermions including
doublers into account. We will discuss the problem in Sec. \ref{s:Sum}.
In Appendix \ref{s:App}, we will give a similar formulation for the one-dimensional 
Su-Schrieffer-Heeger (SSH) model \cite{Su:1979aa},
which may be helpful in understanding the relationship between
the continuum Dirac fermions and the phase of the BBH (or SSH) lattice model.

\section{Dirac fermion model}

In this section, we first introduce the BBH lattice model, and next, take the continuum limit.
The Dirac fermion model thus obtained belongs to class BDI \cite{Altland:1997aa,Schnyder:2008aa}, 
and boundary zero energy states can be easily obtained.

The BBH model introduced in \cite{Benalcazar:2017ab,Benalcazar:2017aa} is a two-dimensional version 
of the SSH model \cite{Su:1979aa}.
The BBH Hamiltonian is defined explicitly by
\begin{alignat}1
H(\bm k)&=\Gamma^j g_j(\bm k),
\label{BBHHam1}
\end{alignat}
where $g_j(k)$ is given by
\begin{alignat}1
&g_1=\lambda_x\sin k_x
\nonumber\\
&g_2=\lambda_y\sin k_y
\nonumber\\
&g_3=\gamma_x+\lambda_x \cos k_x
\nonumber\\
&g_4=\gamma_y+\lambda_y \cos k_y.
\label{BBHHam2}
\end{alignat}
Here, $\gamma_j$ is the hopping within a unit cell, whereas $\lambda_j$ is the hopping between the unit cells in the $j=x,y$ direction.
Benalcazar {\it et al.} have chosen the $\Gamma$-matrices such that
$\Gamma^1=-\sigma^3\otimes\sigma^2$, $\Gamma^2=-\sigma^1\otimes\sigma^2$, $\Gamma^3=1\otimes\sigma^1$, 
and $\Gamma^4=-\sigma^2\otimes\sigma^2$ as well as $\Gamma_5=-1\otimes\sigma^3$, 
but any other definitions may be possible if they obey
$\{\Gamma^j,\Gamma^l\}=2\delta^{jl}$ ($j,l=1,\cdots,4$) 
and $\Gamma_5=(-i)^2\Gamma^1\Gamma^2\Gamma^3\Gamma^4$, so that
$\tr\Gamma_5\Gamma^1\Gamma^2\Gamma^3\Gamma^4=(2i)^2 $.

\subsection{Continuum limit of the BBH model}

The lattice model (\ref{BBHHam2}) includes four Dirac fermions at $k_j=0,\pi$. 
\begin{alignat}1
H_{(0,0)}&=
+\Gamma^1\lambda_xk_1+\Gamma^2\lambda_yk_2+\Gamma^3(\gamma_x+\lambda_x)+\Gamma^4(\gamma_y+\lambda_y),
\nonumber\\
H_{(\pi,0)}&
=-\Gamma^1\lambda_xk_1+\Gamma^2\lambda_yk_2+\Gamma^3(\gamma_x-\lambda_x)+\Gamma^4(\gamma_y+\lambda_y),
\nonumber\\
H_{(0,\pi)}&
=+\Gamma^1\lambda_xk_1-\Gamma^2\lambda_yk_2+\Gamma^3(\gamma_x+\lambda_x)+\Gamma^4(\gamma_y-\lambda_y),
\nonumber\\
H_{(\pi,\pi)}&
=-\Gamma^1\lambda_xk_1-\Gamma^2\lambda_yk_2+\Gamma^3(\gamma_x-\lambda_x)+\Gamma^4(\gamma_y-\lambda_y).
\label{BBHDirFou}
\end{alignat}
As we will argue in Sec. \ref{s:Sum}, we need to consider all the contributions from these Dirac fermions to clarify the topological 
property of the lattice BBH model. However, for the time being, we consider a single Dirac fermion of the form,
\begin{alignat}1
H(\bm k)=\Gamma^jk_j+\Gamma^{a+2}\phi_a,
\label{DirBBH}
\end{alignat}
where $j=1,2$ ($x^1=x, x^2=y$), $a=1,2$.
This is a model of the two-dimensional Dirac fermion coupled with a O(2) Higgs field $\bm\phi=(\phi_1,\phi_2)$,  
belonging to class BDI with time-reversal, particle-hole, and chiral symmetries
denoted, respectively,  by
\begin{alignat}1
&TH(\bm k)T^{-1}=H(-\bm k) ,
\nonumber\\
&CH(\bm k)C^{-1}=-H(-\bm k) ,
\nonumber\\
&\Gamma_5H(\bm k)\Gamma_5^{-1}=-H(\bm k),
\label{Sym}
\end{alignat}
where $T=K$ (taking complex conjugate) and $C=\Gamma_5K$.
In addition to these, this Hamiltonian has reflection symmetries.
\begin{alignat}1
&M_xH(k_x,k_y)M_x^{-1}=H(-k_x,k_y),
\nonumber\\
&M_yH(k_x,k_y)M_y^{-1}=H(k_x,-k_y),
\label{RefSym}
\end{alignat}
where 
$M_x=i\Gamma^1\Gamma_5$ and
$M_y=i\Gamma^2\Gamma_5$.

\subsection{Boundary zero energy states}

\begin{figure}[htb]
\begin{center}
\includegraphics[scale=0.45]{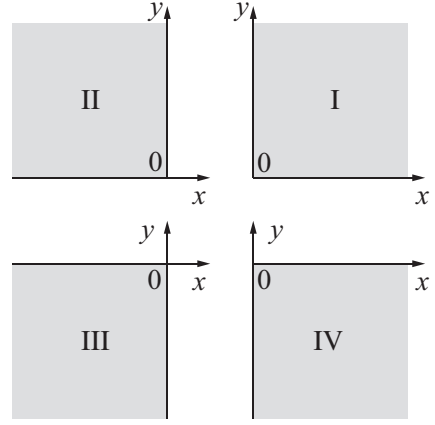}
\caption{
Four types of corners.
}
\label{f:corner}
\end{center}
\end{figure}

The energy eigenvalues of the Hamiltonian (\ref{DirBBH}) are gapped at the zero energy,  
given by $E=\pm\sqrt{\sin^2 k_j+\phi_j^2}$.
However, if the system has boundaries, the model allows zero energy states, $H\psi=0$.
The Hamiltonian in the coordinate representation is given by
\begin{alignat}1
H&=\left(
\begin{array}{cc}
 &D^\dagger \\
D&
\end{array}
\right),
\nonumber\\
D&=-\sigma^3\partial_x-\sigma^1\partial_y-i\sigma^2\phi_y+\phi_x
\nonumber\\
&=\left(\begin{array}{rr}-\partial_x+\phi_x&-\partial_y-\phi_y\\-\partial_y+\phi_y&\partial_x+\phi_x\end{array}\right).
\label{DirHamCor}
\end{alignat}
For simplicity, we here assume $\phi_x$ and $\phi_y$ are positive constants, $\phi_x, \phi_y>0$. 
Let us set $\psi=(\xi,\eta)^T$, where $\xi$ and $\eta$ is the wave function with chirality $\Gamma_5=-1$ and $1$, respectively. 
Then, the zero energy equation reads
\begin{alignat}1
D\xi=0, \quad D^\dagger \eta=0,
\end{alignat}
We readily obtain the following normalizable solution depending on the boundaries:
\begin{alignat}1
&\xi_{\rm I}={\cal N}\left(\begin{array}{c}0\\e^{-\phi_xx-\phi_yy}\end{array}\right),
\eta_{\rm I}=0,
\quad (x>0, y>0)
\label{Cor1}
\\
&\xi_{\rm III}={\cal N}\left(\begin{array}{c}e^{\phi_xx+\phi_yy}\\0\end{array}\right),
\eta_{\rm III}=0,
\quad (x<0, y<0)
\label{Cor2}
\\
&
\xi_{\rm II}=0,
\eta_{\rm II}={\cal N}\left(\begin{array}{c}e^{\phi_xx-\phi_yy}\\0\end{array}\right),
\quad (x<0, y>0)
\label{Cor3}
\\
&
\xi_{\rm IV}=0,
\eta_{\rm IV}={\cal N}\left(\begin{array}{c}0\\e^{-\phi_xx+\phi_yy}\end{array}\right),
\quad (x>0, y<0)
\label{Cor4}
\end{alignat}
where the normalization constant ${\cal N}=2\sqrt{\phi_x\phi_y}$.
In the limit $\phi_x,\phi_y\rightarrow +\infty$, the above wave functions become $\xi^2$ or $\eta^2\rightarrow\delta(x)\delta(y)$, localized at 
the origin.
Thus, it turns out that the present model allows corner states rather than conventional edge states.
See Appendix J in \cite{Benalcazar:2017aa}.

\subsection{Symmetry-breaking perturbations}
\label{s:sym_breaking_BBH}
So far we have derived the Dirac Hamiltonians of the type (\ref{DirBBH}) and its
corner states Eqs. (\ref{Cor1})-(\ref{Cor4}).
It turns out that the two independent mass terms protected by reflection symmetries
are responsible for the corner states.
However, reflection symmetries allow another mass term given by
\begin{alignat}1
&H_{\rm sb} = i\Gamma^3\Gamma^4 \phi_{12},
\label{SymBrePer}
\end{alignat}
which has broken chiral and time reversal symmetries
(but unbroken particle-hole symmetry).
Therefore, if we require auxiliary time reversal symmetry in Eq. (\ref{Sym}) as well as reflection symmetries Eq. (\ref{RefSym}),
the Dirac Hamiltonian  (\ref{DirBBH}) has inevitably chiral symmetry.
This is similar to the SSH Dirac model (\ref{SinDirHam}), 
in which one of symmetries, e.g., inversion symmetry results in time reversal,
chiral symmetries, etc. 
Such symmetry properties 
may be due to the fact that the Dirac models (\ref{DirBBH}) and (\ref{SinDirHam}) are minimal models 
with a corner state and an edge state, respectively.

In the next section,  we will investigate the topological properties of the edge states 
without the mass term (\ref{SymBrePer}).
It will also turn out that even with Eq. (\ref{SymBrePer}), 
the corner state is still protected by particle-hole symmetry,  as will be presented in Sec. \ref{s:Sum}.

Another simple way to break chiral symmetry (as well as particle-hole symmetry) is to extend the model to the layer systems. 
In Appendix \ref{s:sym_breaking}, we will demonstrate a similar extension of the SSH model to
ladder models. We will show that 
although the lattice model does not have chiral symmetry,
the Dirac fermions with chiral symmetry are responsible for the existence of
the edge states and hence the index theorem is a useful tool to investigate them.
Even for the present BBH model, such an argument can also be applied, since the Hamiltonian of a layered 
BBH model is obtained by replacing $H_{\rm SSH}$ and  $H_{\rm SSH}'$ with  $H_{\rm BBH}$ and  $H_{\rm BBH}'$
in Eq. (\ref{SSHLad}) .
This implies  that the Dirac fermion
(\ref{DirBBH})  is the basic effective model describing the corner states.
In particular, the corner states of the trilayer system, which has broken chiral symmetry, can be described 
by the Dirac fermion (\ref{DirBBH}) near half-filling.


\section{Index theorem}
\label{s:index}

The corner states has a topological origin which is the same as the Jackiw-Rossi states localized at a point defect (vortex)
given by $\bm\phi=\Delta(r)(\cos\theta,\sin\theta)$ \cite{JackiwRossi:1981}. 
To see this, we will consider  smooth boundaries introduced by a coordinate-dependent Higgs field such that
\begin{alignat}1
\phi_1=\phi_x(\bm x),\quad \phi_2=\phi_y(\bm y),
\label{Hig}
\end{alignat}
where  $\phi_x$ and $\phi_y$ depends generically on $\bm x$ with the asymptotic form
$\phi_x(x\rightarrow\pm\infty,y)\rightarrow \mbox{const. }$ and $\phi_y(x,y\rightarrow\pm\infty)\rightarrow \mbox{const}$,
respectively.
For such a model, we will apply the index theorem on open spaces 
\cite{Callias,Weinberg:1981uq,Niemi:1984aa,niemisemenoff86R,Fukui:2010aa}.
We assume that the dependence of $\phi_j$ on $x^j$ is so smooth that  $|\partial_j\phi_j|\ll |\phi_j|$ is valid, implying that 
we can make use of  the derivative expansion in the following calculations.

\begin{figure}[htb]
\begin{center}
\includegraphics[scale=0.5]{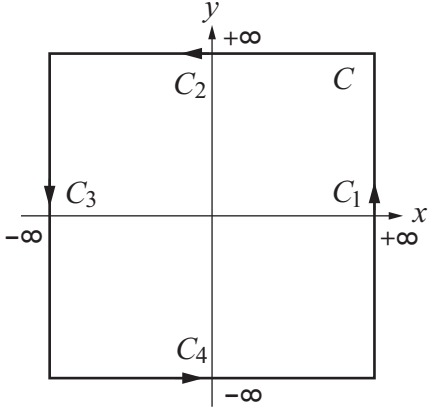}
\caption{
The integration contour $C$ of $\langle j^j_5(x)\rangle$ in Eq. (\ref{IndThe}).
The closed path $C$ is divided into four lines denoted by $C_j$ ($j=1,\cdots,4$).
}
\label{f:contour}
\end{center}
\end{figure}

Since the model has chiral symmetry, the zero energy states can be labeled by the chirality.
Therefore, let us define the index of $H$ such that
\begin{alignat}1
{\rm ind }~H&=n_+-n_-
\nonumber\\
&=\lim_{m\rightarrow0}{\rm Tr }~\Gamma_5\frac{m^2}{H^2+m^2},
\label{IndTheOri}
\end{alignat}
where $n_\pm$ stands for the number of zero energy states with chirality $\pm1$.
As we will show below, the rhs of the above index can be expressed by the axial vector current
\begin{alignat}1
\langle j_5^j(\bm x)\rangle=\lim_{\bm y\rightarrow \bm x}{\rm tr ~}\Gamma_5\Gamma^j\Big(\frac{1}{iH+m}-\frac{1}{iH+M}\Big)
\delta(\bm x-\bm y),
\end{alignat}
where the second term with $M$ is a Pauli-Villars regulator.
In the following calculation, the regulator will be suppressed.
The divergence of the current yields
\begin{alignat}1
\partial_j\langle j_5^j\rangle
&=\partial_j^x\lim_{\bm y\rightarrow \bm x}{\rm tr}\,\Gamma_5\Gamma^j \frac{1}{iH+m}\delta(\bm x-\bm y)
\nonumber\\
&=\lim_{\bm y\rightarrow \bm x}{\rm tr ~}\Gamma_5\Gamma^j(\partial_j^x+\partial_j^{y})
\frac{1}{iH+m}\delta(\bm x-\bm y)
\nonumber\\
&=\lim_{\bm y\rightarrow \bm x}{\rm tr}\,\Gamma_5
\Big(\Gamma^j\partial_j^x\frac{1}{iH+m}+\frac{1}{iH+m}\Gamma^j\partial_j^x\Big)\delta(\bm x-\bm y).
\end{alignat}
Here, note that $iH=\Gamma^j\partial_j+i\Gamma^{a+2}\phi_a$, and 
the Higgs term anti-commutes with $\Gamma_5$. Then, we have
\begin{alignat}1
\partial_j\langle j_5^j\rangle
&=\lim_{\bm y\rightarrow \bm x}{\rm tr }\,\Gamma_5(2iH)\frac{1}{iH+m}\delta(\bm x-\bm y)
\nonumber\\
&=2\lim_{\bm y\rightarrow \bm x}{\rm tr }\,\Gamma_5\Big(1-\frac{m}{iH+m}\Big)\delta(\bm x-\bm y)
\end{alignat}
The term $1$ in the parentheses cancels the same one in the regulator.
Thus, we have
\begin{alignat}1
\partial_j\langle j_5^j\rangle
&=-2\lim_{\bm y\rightarrow \bm x}{\rm tr }\,\Gamma_5\frac{m(-iH+m)}{H^2+m^2}\delta(\bm x-\bm y).
\end{alignat}
Since $H$ anti-commutes with $\Gamma_5$, we reach
\begin{alignat}1
\partial_j\langle j_5^j\rangle
=&-2\lim_{\bm y\rightarrow \bm x}{\rm tr }\,\Gamma_5\frac{m^2}{H^2+m^2}\delta(\bm x-\bm y)
\nonumber\\
&+2\lim_{\bm y\rightarrow \bm x}{\rm tr }\,\Gamma_5\frac{M^2}{H^2+M^2}\delta(\bm x-\bm y),
\end{alignat}
where we have explicitly denoted the contribution from the regulator.
Thus, integrating over the space and taking the limit $m\rightarrow0$ and $M\rightarrow\infty$ yields
\begin{alignat}1
{\rm ind }~H&=
-\frac{1}{2}\int \partial_j\langle j_5^j\rangle d^2x+c_1
\nonumber\\
&=
-\frac{1}{2}\oint_{C} \epsilon_{jl}\langle j_5^j\rangle dx^l+c_1,
\label{IndThe}
\end{alignat}
where $C$ is the contour denoted in Fig. \ref{f:contour}, and
\begin{alignat}1
c_1=\lim_{M\rightarrow \infty}{\rm Tr }\,\Gamma_5\frac{M^2}{H^2+M^2},
\end{alignat}
is associated with the chiral anomaly.
Although this vanishes trivially in the present case with no gauge potentials, it plays an important role 
in the reproduction of the correct index for the Jackiw-Rossi model 
in a magnetic field \cite{Weinberg:1981uq,Fujiwara:2012aa}.

Let us now compute the current in the derivative expansion:
\begin{alignat}1
\langle j_5^j(\bm x)\rangle&=\lim_{\bm y\rightarrow \bm x}{\rm tr }\,\Gamma_5\Gamma^j\frac{1}{iH+m}\delta(\bm x-\bm y)
\nonumber\\
&=\lim_{\bm y\rightarrow \bm x}\int\frac{d^2k}{(2\pi)^2}{\rm tr }\,\Gamma_5\Gamma^j\frac{1}{iH+m}e^{i\bm k\cdot(\bm x-\bm y)}
\nonumber\\
&=\int\frac{d^2k}{(2\pi)^2}{\rm tr }\,\Gamma_5\Gamma^j e^{-i\bm k\cdot\bm x}\frac{-iH+m}{H^2+m^2}e^{i\bm k\cdot\bm x},
\label{AxiCur}
\end{alignat}
where the Pauli-Villars regulator has been suppressed.
Note here that
\begin{alignat}1
&e^{-i\bm k\cdot\bm x}He^{i\bm k\cdot\bm x}=\Gamma^j k_j+\Gamma^{a+2}\phi_a+O_1,
\nonumber\\
&e^{-i\bm k\cdot\bm x}H^2e^{i\bm k\cdot\bm x}=\bm k^2+\phi^2-i\Gamma^j\Gamma^{a+2}\partial_j\phi_a+O_2,
\label{PlaWavHam}
\end{alignat}
where $\phi^2=\phi_a^2$, $O_1=-i\Gamma^\mu\partial_\mu$, and $O_2=-2ik^\mu\partial_\mu-\partial_\mu^2$.

The contour integration in Eq. (\ref{IndThe}) can be carried out by  dividing the path $C$ into four lines $C_j$.
To compute the integration on the line $C_1$, 
let us consider the limit
$x\rightarrow\infty$ and regard $\phi_x=\phi_x(+\infty)$ as a constant. Then, $\langle j_5^1(x)\rangle$ at $x^1\rightarrow+\infty$
can be calculated as follows:
\begin{alignat}1
\langle j_5^1&(x^1\rightarrow+\infty)\rangle=\int\frac{d^2k}{(2\pi)^2}{\rm tr }\,\Gamma_5\Gamma^1
\nonumber\\
&\times\frac{-\Gamma^j(\partial_j+i k_j)-i\Gamma^{3}\phi_x(\infty)-i\Gamma^4\phi_2+O_1}{\bm k^2+\phi^2
-i\Gamma^2\Gamma^4\partial_2\phi_2+O_2},
\label{DerExp}
\end{alignat}
where we have safely taken the limit $m\rightarrow0$, while the regulator has vanished in the limit $M\rightarrow\infty$.
Note that in the denominator above we assume $|\partial_2\phi_2|\ll \phi^2$.  Therefore, as the leading contribution,
we have
\begin{alignat}1
\langle j_5^1\rangle&=\int\frac{d^2k}{(2\pi)^2}{\rm tr }\,\Gamma_5\Gamma^1
\big(-i\Gamma^{3}\phi_x(\infty)\big)
\frac{i\Gamma^2\Gamma^4\partial_2\phi_2}{(\bm k^2+\phi^2)^2}
\nonumber\\
&=
\frac{1}{\pi}\frac{\phi_x(\infty)\partial_2\phi_2}{\phi_x(\infty)^2+\phi_2^2}.
\end{alignat}
Thus, the integration on $C_1$ yields
\begin{alignat}1
\int_{C_1} \langle j_5^1(x)\rangle dy
&=\int_{-\infty}^\infty \langle j_5^1(x^1=\infty)\rangle dy
\nonumber\\
&=\frac{1}{\pi}
\Bigg[\arctan\frac{\phi_y(\infty)}{\phi_x(\infty)}-\arctan\frac{\phi_y(-\infty)}{\phi_x(\infty)}
\Bigg].
\end{alignat}
Integration on the other lines $C_j$ can be computed likewise, and we finally reach
\begin{widetext}
\begin{alignat}1
\int_{C} \epsilon_{jl}\langle j_5^j\rangle dx^l
=\frac{1}{\pi}
\Bigg[&
\arctan\frac{\phi_y(\infty)}{\phi_x(\infty)}-\arctan\frac{\phi_y(-\infty)}{\phi_x(\infty)}
+\arctan\frac{\phi_x(\infty)}{\phi_y(\infty)}-\arctan\frac{\phi_x(-\infty)}{\phi_y(\infty)}
\nonumber\\
&
-\arctan\frac{\phi_y(\infty)}{\phi_x(-\infty)}+\arctan\frac{\phi_y(-\infty)}{\phi_x(-\infty)}
-\arctan\frac{\phi_x(\infty)}{\phi_y(-\infty)}+\arctan\frac{\phi_x(-\infty)}{\phi_y(-\infty)}
\Bigg]
\nonumber\\
=\frac{1}{2}\Bigg[&{\rm sgn} \frac{\phi_y(\infty)}{\phi_x(\infty)}+{\rm sgn} \frac{\phi_y(-\infty)}{\phi_x(-\infty)}
-{\rm sgn} \frac{\phi_y(-\infty)}{\phi_x(\infty)}-{\rm sgn} \frac{\phi_y(\infty)}{\phi_x(-\infty)}
\Bigg],
\end{alignat}
where we have used the identity $\arctan x+\arctan x^{-1}=\frac{\pi}{2}{\rm sgn}\, x$.
It thus turns out that the index is given by
\begin{alignat}1
{\rm ind }~H
&=-\frac{1}{4}\Bigg[{\rm sgn} \frac{\phi_y(\infty)}{\phi_x(\infty)}+{\rm sgn} \frac{\phi_y(-\infty)}{\phi_x(-\infty)}
-{\rm sgn} \frac{\phi_y(-\infty)}{\phi_x(\infty)}-{\rm sgn} \frac{\phi_y(\infty)}{\phi_x(-\infty)}
\Bigg].
\label{BBHInd}
\end{alignat} 
\end{widetext}
This is one of the main results in the present paper.
The rhs of the above equation is minus the winding number of $\bm\phi(x)=(\phi_x(x),\phi_y(y))$ around $C$:
If $\phi_x(-\infty)<0<\phi_x(+\infty)$, and $\phi_y(-\infty)<0<\phi_y(+\infty)$, the winding number equals 1,
and the index of $H$ is given by $-1$. This configuration of $\bm\phi$ corresponds to Eq. (\ref{Cor1})
which has chirality $-1$. Other cases in Eqs.(\ref{Cor2})-(\ref{Cor4}) match the index given in Eq. (\ref{BBHInd}).
For example, the corner state (\ref{Cor3}) can be realized by $\phi_x(+\infty)<0<\phi_x(-\infty)$ and $\phi_y(-\infty)<0<\phi_y(+\infty)$,
which has winding number $-1$, and hence ind $H$=1.
Therefore, it turns out that a single corner  can be regarded as a point defect, and
the zero energy state of the BBH Dirac fermion is the same class as the Jackiw-Rossi states localized in a vortex.

In passing, we mention that if the Hamiltonian includes vector and/or axial vector gauge potentials, 
the boundary integration in Eq. (\ref{IndThe}) needs careful calculations, constructing the boundary operators and
computing their spectral flow \cite{niemisemenoff86R,Shiozaki:2012aa}.
In the present case, however, the model is simple enough to reproduce the index by the simple derivative expansion. 

\section{Dipole pump}
\label{s:pump}

Benalcazar {\it et al.} have proposed a dipole pump and demonstrated it for the BBH model \cite{Benalcazar:2017aa}.
The continuum Dirac model presented in this paper corresponds to the case with a single corner, which can be realized 
by a coordinate-dependent Higgs field introduced in the previous section.
In this section, we further introduce a symmetry-breaking term dependent on time  and calculate the vector current density to investigate a
charge pump associated with a corner.
Calculations of this section is parallel to those in Ref. \cite{Fukui:2017aa}.

As discussed by BBH, we consider the model which includes symmetry breaking (reflection symmetries, in particular) term 
\begin{alignat}1
H=-i\Gamma^j\partial_j+\sum_{a=1,2}\Gamma^{a+2}\phi_a+\Gamma_5\phi_3,
\label{HamPum}
\end{alignat}
where we assume that $\phi_a$ $(a=1,2,3)$ depends not only $\bm x$ but also $t$, $\phi_a=\phi_a(t,\bm x)$.
For the time being, we only assume that $|\partial_\mu\phi_a|\ll |\phi|$, which allows the derivative expansion.
The Lagrangian corresponding to the Hamiltonian (\ref{HamPum}) is
\begin{alignat}1
{\cal L}&=\bar\psi(i\gamma^\mu\partial_\mu-\gamma^3\phi_1-\phi_2-i\gamma_5\phi_3)\psi
\nonumber\\
&\equiv
\bar\psi(i\slashed\partial-\Phi-\phi_2)\psi,
\end{alignat}
where $\mu=0,1,2$  $(x^0=t)$, and 
$\gamma$-matrices obeying $\{\gamma^\mu,\gamma^\nu\}=2g^{\mu\nu}$ with $g^{\mu\nu}=\mbox{diag}(1,-1,-1,-1)$ is defined by
$\gamma^0=\Gamma^4$, $\gamma^j=\gamma^0\Gamma^j$ ($j=1,2$), $\gamma^3=\gamma^0\Gamma^3$, and
$\gamma_5=i\gamma^0\gamma^1\gamma^2\gamma^3$.
The U(1) vector current is defined by
\begin{alignat}1
\langle j^\mu(x)\rangle&=\langle0|\bar\psi(x)\gamma^\mu\psi(x)|0\rangle
\nonumber\\
&=-\lim_{y\rightarrow x}\langle0|T\gamma^\mu\psi(x)\bar\psi(y)|0\rangle,
\end{alignat}
where $x=(t,\bm x)$ and  the propagator is given by
\begin{alignat}1
\langle0|T\psi(x)\bar\psi(y)|0\rangle
=
\frac{i}{i\slashed\partial-\Phi-\phi_2+i\epsilon}\delta(x-y)
\end{alignat}
In the plane wave representation of the delta function similar to Eq. (\ref{AxiCur}),
we have
\begin{alignat}1
\langle j^\mu(x)\rangle=&
\int\frac{d^3k}{i(2\pi)^3}e^{-ikx}
\tr \gamma^\mu\frac{1}{i\slashed\partial-\Phi-\phi_2+i\epsilon}e^{ikx},
\end{alignat}
where $kx=\omega t-\bm k\cdot\bm x$.
Using
\begin{alignat}1
\big((i\slashed\partial-\Phi)&-\phi_2\big)\big(-(i\slashed\partial-\Phi)-\phi_2\big)
\nonumber\\
&=\partial^2+\phi^2+i(\slashed\partial\Phi)-i(\slashed\partial\phi_2),
\end{alignat}
we reach
\begin{alignat}1
\langle j^\mu(x)\rangle=&
\int\frac{d^3k}{i(2\pi)^3}
\tr \gamma^\mu
(-i\slashed\partial+\slashed k+\Phi-\phi_2)
\nonumber\\
&\times\frac{1}{(\partial+ik)^2+\phi^2+i(\slashed\partial\Phi)-i(\slashed\partial\phi_2)-i\epsilon},
\end{alignat}
where $\phi^2=\sum_{a=1}^3\phi_a^2$.
Since we assume that the $x$ dependence of $\phi_j(x)$ is so smooth that the derivative expansion is a good approximation, 
as has done in Sec. \ref{s:index},
the leading contribution is
\begin{alignat}1
\langle j^\mu(x)\rangle
=&\int\frac{d^3k}{i(2\pi)^3}
\tr \gamma^\mu
\frac{(\Phi-\phi_2)\big[i(\slashed\partial\Phi)-i(\slashed\partial\phi_2)\big]^2}{(\phi^2-k^2-i\epsilon)^3}
\nonumber\\
&+O\big((\partial\phi)^{-3}\big),
\end{alignat}
where $k^2=\omega^2-\bm k^2$ and $\phi^2=\sum_{a=1}^3\phi_a^2$.
Note that
\begin{alignat}1
\tr \gamma^\mu&(\Phi-\phi_2)\big[i(\slashed\partial\Phi)-i(\slashed\partial\phi_2)\big]^2
\nonumber\\
=&\tr\gamma^\mu(\gamma^3\phi_1+i\gamma_5\phi_3-\phi_2)
\nonumber\\
&\times\big[i\gamma^\nu\gamma^3\partial_\nu\phi_1-\gamma^\nu\gamma_5\partial_\nu\phi_3
-i\gamma^\nu\partial_\nu\phi_2)\big]^2
\nonumber\\
=&4\epsilon^{\mu\nu\rho}\epsilon^{abc}\phi_a\partial_\nu\phi_b\partial_\rho\phi_c,
\end{alignat}
where we have used $\tr\gamma_5\gamma^\mu\gamma^\nu\gamma^\rho\gamma^\sigma=-4i\epsilon^{\mu\nu\rho\sigma}$.
Thus, we reach
\begin{alignat}1
\langle j^\mu(x)\rangle
&=4\epsilon^{\mu\nu\rho}\epsilon^{abc}\phi_a\partial_\nu\phi_b\partial_\rho\phi_c\int\frac{d^3k}{i(2\pi)^3}
\frac{1}{(\phi^2-k^2-i\epsilon)^3}
\nonumber\\
&=\frac{1}{8\pi \phi^3}
\epsilon^{\mu\nu\rho}\epsilon^{abc}\phi_a\partial_\nu\phi_b\partial_\rho\phi_c
\nonumber\\
&=\frac{1}{8\pi}
\epsilon^{\mu\nu\rho}\epsilon^{abc}\hat\phi_a\partial_\nu\hat\phi_b\partial_\rho\hat\phi_c,
\end{alignat}
where $\hat\phi_a\equiv \phi_a/\phi$.
This is another main result in this paper.
The above current density is indeed topological: The total pumped charge is given by the integration of the current density over 
S$^2$ embedded in 2+1 dimensions, which is given by the winding number of the Higgs field on S$^2$. 

To be more specific, let us compute the total current 
by integrating over $-\infty<x^0(=t)<\infty$ as well as over $C$ in Fig. \ref{f:contour}. 
To this end, we specify $\phi_a$ such that 
\begin{alignat}1
&\phi_1(x)=\phi_x(x^1)\sin \theta(x^0),
\nonumber\\
&\phi_2(x)=\phi_y(x^2)\sin \theta(x^0),
\nonumber\\
&\phi_3(x)=\phi_0\cos \theta(x^0),
\end{alignat}
where $\phi_{x,y}$ is basically the same as Eq. (\ref{Hig}), and $\phi_0$ is a constant. We further assume 
the asymptotic behavior of the Higgs field as
$|\phi_x(x^1\rightarrow\pm\infty)|\rightarrow|\phi_0|$  and $|\phi_y(x^2\rightarrow\pm\infty)|\rightarrow|\phi_0|$ for the spatial part,
whereas $\theta(x^0\rightarrow-\infty)=0$ and $\theta(x^0\rightarrow+\infty)=\pi$ for the temporal part. This implies that
a trivial system at $t=-\infty$ becomes another trivial system at $t=+\infty$ via the topological state studied in Sec. \ref{s:index}.

We compute $j^1(x)$ on $C_1$.
At $x^1\rightarrow+\infty$, we have
$\phi^2=\phi_0^2+\phi_y^2(x^2)\sin^2\theta(x^0)$. Therefore, 
\begin{alignat}1
\langle j^1(x^1\rightarrow\infty)\rangle
&=\frac{2\epsilon^{120}\epsilon^{abc}\phi_a\partial_2\phi_b\partial_0\phi_c}{8\pi (\phi_0^2+\phi_y^2\sin^2\theta)^\frac{3}{2}}
\nonumber\\
&=-\frac{\phi_x(\infty)\partial_2\phi_y\sin\theta\partial_0\theta}{4\pi (\phi_0^2+\phi_y^2\sin^2\theta)^\frac{3}{2}}.
\end{alignat}
Thus, integration over $x^0(=t)$ gives
\begin{alignat}1
\int_{-\infty}^{\infty}\langle j^1(x^1=\infty)\rangle dx^0
&=\int_{0}^{\pi}\langle j^1(x^1=\infty)\rangle d\theta
\nonumber\\
&=\frac{-\phi_x(\infty)}{2\pi\phi_0}\frac{\partial_y\phi_y(y)}{\phi_y^2(y)+\phi_0^2}.
\end{alignat}
Furthermore, the integration of the above on $C_1$ gives
\begin{alignat}1
\int_{C_1}dx^2&\int_{-\infty}^{\infty}dx^0\langle j^1(x)\rangle
\nonumber\\
&=\frac{-1}{2\pi}\Big[
\arctan\frac{\phi_y(\infty)}{\phi_x(\infty)}-\arctan\frac{\phi_y(-\infty)}{\phi_x(\infty)}
\Big].
\end{alignat}
Together with the contributions on the  other lines $C_j$, the total pumped charge $Q$ is just the index of $H$ in Eq. (\ref{BBHInd}),
$
|Q|=|\mbox{ind}\,H| 
$, where the sign is generically dependent on the process of the pump.
This implies that during the adiabatic change of the ground state between 
the two trivial ground states with $\bm \phi=(0,0,\pm\phi_0)$ via topological one with $\bm \phi=(\phi_x,\phi_y,0)$, 
the integral charge, just corresponding to the index
(or roughly speaking, the number of the zero energy states) of the Hamiltonian 
with $\bm \phi=(\phi_x,\phi_y,0)$, flows, which is expected in the topological pump.

\section{Summary and discussion}
\label{s:Sum}

In summary, we studied the topological aspects of the Dirac fermion coupled with a two-component Higgs field, which is
a naive continuum limit of the BBH lattice model.
Since this model has chiral symmetry, we first investigated the zero energy corner states using the index theorem in Sec. \ref{s:index}.
We argued that a corner can be regarded as a point defect, and hence, 
the corner states are in the same class of the Jackiw-Rossi states localized at a vortex.

Generically, the Dirac fermion $H=-i\sigma^j\partial_j+\sigma^{d+1}\phi$ in $d=1,2$ dimensions
is topological \cite{Jackiw:1976fk,JackiwRossi:1981,Goldstone:1981aa}
in the sense that it has Berry phase $\pm\pi/2$ in $d=1$ 
(See Sec. \ref{s:single})
and
Chern number $\pm1/2$ in $d=2$. 
This is why the zero energy state appears in these Dirac fermions.
In a similar reason, we  can tell that the present Dirac fermion has a corner state because  
the Higgs field $\bm\phi$ has nontrivial winding number, as we have shown in this paper.

\begin{table}[h]
\begin{center}
\begin{tabular}{c|cccc}
\hline
$\alpha$ & $(0,0)$& $(\pi,0)$& $(0,\pi)$& $(\pi,\pi)$\\
\hline\hline
$\phi_x$& $1+\gamma_x$ & $1-\gamma_x$ & $1+\gamma_x$ & $1-\gamma_x$\\
$\phi_y$& $1+\gamma_y$ &$1+\gamma_y$ & $1-\gamma_y$& $1-\gamma_y$\\
\hline
\end{tabular}
\end{center}
\caption{
Higgs parameters $\bm\phi_{\alpha}$ at four points $\alpha$. 
We have set $\lambda_x=\lambda_y=1$.
}
\label{t:Higgs}
\end{table}

On the other hand,  when we argue the quadrupole phase of the lattice BBH model
in terms of the Dirac fermions, 
we need to take account of the four Dirac fermions derived in Eq. (\ref{BBHDirFou}).
See also Appendix \ref{s:App}, where we discuss the same problem for  the SSH model.
After changing the sign of the two or four $\Gamma$ matrices associated with $k_j=\pi$ as demonstrated in Eq. 
(\ref{SinDirHam}), we have 
\begin{alignat}1
H_\alpha=\Gamma^jk_j+\Gamma^{a+2}\phi_{\alpha,a},
\label{EacDir}
\end{alignat}
where $\alpha=(0,0),(\pi,0),(0,\pi),(\pi,\pi)$, and $\bm\phi_\alpha$ is summarized in Table \ref{t:Higgs},
where we have set $\lambda_x=\lambda_y=1$.
From Eq. (\ref{BBHInd}), 
the topological invariant to characterize each Dirac fermion may be assigned in a similar way demonstrated in
Appendix \ref{s:single} such that
\begin{alignat}1
Q_\alpha&=\frac{1}{4}{\rm sgn\,}\frac{\phi_{\alpha,y}}{\phi_{\alpha,x}}=
\frac{1}{4}({\rm sgn\,}\phi_{\alpha,x})({\rm sgn\,}\phi_{\alpha,y})
\nonumber\\
&=\frac{q_{\alpha,x}}{\pi}\frac{q_{\alpha,y}}{\pi},
\label{EacTopInv}
\end{alignat}
where $q$ is the Berry phase (with a specific gauge) for one-dimensional Dirac fermion defined in Eq. (\ref{IndBer}).
According to the same discussion in Eqs. (\ref{WavFun}) and (\ref{LatWavFun}),
the total topological invariant for the lattice BBH model is the sum of all $Q_\alpha$,
\begin{alignat}1
Q_{\rm BBH}=\sum_\alpha Q_\alpha.
\end{alignat}
Form Table \ref{t:Higgs} we arrive at $Q_{\rm BBH}=1$
only when $|\gamma_x|<1(=\lambda_x)$ and $|\gamma_y|<1(=\lambda_y)$.

It should be noted that the Dirac fermion (\ref{EacDir}) can be characterized by 
(\ref{EacTopInv}) which is given by the Berry phases for $x$ and $y$ directions.
Indeed, the lattice BBH model has been characterized by  
the Wannier-sector polarization \cite{Benalcazar:2017aa} or entanglement polarization \cite{1805.02831},
which are basically Berry phases of the projected one-dimensional model.

As already mentioned, reflection symmetries as well as time reversal symmetry for two-dimensional Dirac
model allow just two mass terms, which causes chiral symmetry of the Hamiltonian (\ref{DirBBH}). 
This enables us to apply the index theorem to the present model, as carried out in Sec. \ref{s:index}.
Although this is the most general Hamiltonian with reflection symmetries as well as time reversal symmetry,
the model can include the chiral symmetry-breaking mass term $H_{\rm sb}$ (\ref{SymBrePer}),  
if time reversal symmetry is relaxed, as discussed in Sec. \ref{s:sym_breaking_BBH}.
In what follows, we will mention its effect to the corner states.
For the time being, let us consider the case with one corner.
With chiral symmetry, the index theorem tells that 
the Hamiltonian shows at least $q$ zero energy states, when the Higgs field has winding number $q$.
If chiral symmetry
is broken but particle-hole symmetry is unbroken by $H_{\rm sb}$,  
the model shows  $q$ mod 2 zero energy states, since  the zero energy states protected by chiral symmetry 
are lifted pairwise to positive and negative energies. 
Therefore, if $q$ is odd, only one zero energy state is generically protected by particle-hole symmetry
for a system with one corner.
Thus, we conclude that the zero energy corner states are protected by reflection symmetries.

More generic multi-band systems such as a layered BBH model, reflection symmetries allow 
various mass terms with broken chiral and particle-hole symmetries with keeping time reversal symmetry.
Even in such a case, effective Dirac fermion (\ref{DirBBH}) should appear not necessarily at zero energy,
as demonstrated in Appendix \ref{s:App} for the SSH model. 
In a two-leg ladder system introduced in \ref{s:sym_breaking},
the edge states appear at nonzero energies as in Eq. (\ref{SSHLad2}).
Nevertheless, it is obvious that the edge states can be  described by the continuum Dirac fermions (\ref{DouFer}).
In a three-leg ladder system, the band center is indeed the original SSH model, although the whole system has 
broken chiral symmetry. Therefore, it is manifest that the edge states in this case are 
effectively due to the continuum Dirac fermion with chiral symmetry.
This feature is also true for the BBH model.
Therefore, if we tune the Fermi energy, we see that the corner states, if they exist, are due to those of the
Dirac fermion (\ref{DirBBH}), which belongs to the same class of Jackiw-Rossi vortex states.

So far we have discussed the case with only one corner. 
Let us now consider the  system with full open boundary conditions having four corners.
In this case, there can appear four corner states.
Then, reflection symmetries Eq. (\ref{RefSym}) ensure that such four corner states  are fourfold degenerate.
Thus, these states yield the quadrupole charge configuration with an infinitesimal small staggered potential.

This is one of interpretations of the quadrupole phase from the point of view of the continuum Dirac fermion model.
In conclusion, the topological origin of the corner states is attributed to the Jackiw-Rossi states of a Dirac fermion
in the continuum limit, 
and their fourfold degeneracy and resultant quadrupole charge configuration in the lattice BBH model
are guaranteed by the reflection symmetries.


In Sec. \ref{s:pump},
we next introduced a symmetry breaking term  and examined the pump, which corresponds to the 
dipole pump proposed by BBH.
We argued that this pump is also topological and the total pumped charge is the same as the index of the Dirac fermion
studied in Sec. \ref{s:index}.

The dipole pump is realized only if the system has four corners and take the quadrupole charge configuration during the
adiabatic pumping process. We just showed that a single Dirac fermion with a point defect (a smooth corner)
can yield a topological current.
It may be interesting to examine the quadrupole state and 
the dipole pump in terms of four Dirac fermions mentioned above, preferably taking into account
generic BDI symmetry breaking terms.

\section*{Acknowledgments}
We would like to thank Y. Hatsugai, S. Hayashi, K.-I. Imura, T. Misumi and Y. Yoshimura for fruitful discussions.
This work was supported in part by Grants-in-Aid for Scientific Research Numbers 17K05563 and 17H06138
from the Japan Society for the Promotion of Science.


\appendix

\section{A Dirac fermion description of the SSH model}
\label{s:App}

We present in Appendix \ref{s:App}  a simple Dirac fermion description of the one-dimensional SSH model,
which is helpful in understanding the relationship between the continuum Dirac fermions
and the lattice model.

\begin{figure}[htb]
\begin{center}
\includegraphics[scale=0.7]{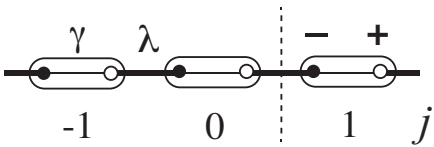}
\caption{
The SSH model on a one-dimensional lattice. An oval including two sites assigned 
the chirality $\pm(=-\sigma^3)$
stands for the unit cell.
The dotted line stands for the boundary discussed in Sec. \ref{s:edge}.
}
\label{f:SSH}
\end{center}
\end{figure}

On a one-dimensional lattice in Fig. \ref{f:SSH}, the SSH model in the momentum representation is defined by
\begin{alignat}1
H_{\rm SSH}(k)=\sigma^2\lambda\sin k+\sigma^1(\gamma+\lambda \cos k),
\label{SSHHam}
\end{alignat}
where $\lambda$ and $\gamma$ are hopping parameters similar to $\gamma_x$ and $\lambda_x$ in Eq. (\ref{BBHHam2}),
respectively.
The crucial symmetry of the model is 
reflection (inversion) symmetry Eq. (\ref{RefSym}) 
implemented by $M_x=\sigma^1$. 
Since the reflection symmetry prohibits a constant term proportional to $\sigma^3$,
the model inevitably has chiral symmetry. 

The above Hamiltonian (\ref{SSHHam})
can be regarded as the Wilson-Dirac Hamiltonian in one dimension, where the first term is 
the kinetic term and the second term is the mass term with the Wilson term ($\sigma^1\lambda\cos k$).
Therefore, in the continuum limit, we have two fermions near $k=0$ and $k=\pi$:
\begin{alignat}1
H_\alpha(k)=\pm\sigma^2k+\sigma^1(\gamma\pm1),\quad \alpha=0,\pi,
\label{DouFer}
\end{alignat}
where we have set $\lambda=1$ for simplicity.
This is due to the doubling mechanics \cite{Nielsen:1981aa,Nielsen:1981ab}.
The continuum Hamiltonian also has reflection symmetry as well as chiral symmetry.

The wave function of the ground state is also expanded around $k\sim0,\pi$ as
\begin{alignat}1
\psi_j&=\int_{-\pi}^\pi \psi(k) e^{ikj}\frac{dk}{2\pi}
\nonumber\\
&\sim\int_{-\Lambda}^{\Lambda} a\psi(k) e^{ikx}\frac{dk}{2\pi}
+\int_{-\Lambda}^{\Lambda} a\psi(k+\frac{\pi}{a}) e^{i(k+\frac{\pi}{a})x}\frac{dk}{2\pi}
\nonumber\\
&\rightarrow\int_{-\infty}^{\infty} 
\left[\psi_0(k) 
+e^{i\frac{\pi}{a}x} \psi_\pi(k) \right]e^{ikx}\frac{dk}{2\pi}
\nonumber\\
&\equiv\psi_0(x)+(-1)^{i\frac{\pi}{a} x}\psi_\pi(x)\equiv\psi(x) ,  
\label{WavFun}
\end{alignat}
where $x\sim aj$ with the lattice constant $a$, 
$\Lambda\sim\pi/a$ is a cutoff parameter and set  $\Lambda\rightarrow\infty$ $(a\rightarrow0)$ in the third line
after the linearization of the dispersion, so the $\psi_0(x)$ and $\psi_\pi(x)$ are slowly varying components.

\subsection{Berry phase}
As we have mentioned, the SSH model includes two fermions described by the Hamiltonian in Eq. (\ref{DouFer}).
The wave function of the lattice model can also be approximated by Eq. (\ref{WavFun}) including the contribution from 
two fermions, so  the Berry connection reads
\begin{alignat}1
\psi^\dagger\partial_k \psi&\sim
\psi_0^\dagger\partial_k\psi_0
+\psi_\pi^\dagger\partial_k\psi_\pi+(\mbox{osc. terms})
\nonumber\\
&\equiv A_0(k)+A_\pi(k),
\label{LatWavFun}
\end{alignat}
where 
the last term is the rapidly oscillating part of the Berry connection including 
$(-)^{i\frac{\pi}{a}x}$ which can be neglected.
Thus, for the lattice model,
the Berry phase
could be given by the sum of  those of the two Dirac fermions,
\begin{alignat}1
q_{\rm SSH}=q_0+q_\pi.
\label{SSHBerPha}
\end{alignat}
where 
\begin{alignat}1
q_\alpha=\frac{1}{i}\int_{-\infty}^{\infty}A_\alpha(k)dk.
\label{DirBerPha}
\end{alignat}

\begin{figure}[htb]
\begin{center}
\includegraphics[scale=0.5]{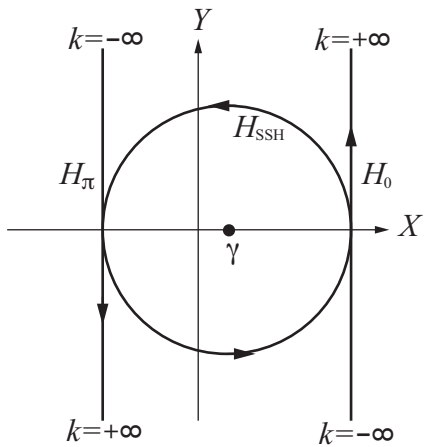}
\caption{
The integration path for the Berry phase of the SSH model Eq. (\ref{SSHHam}) 
(circle)
as we  as of its continuum limit Eq. (\ref{DouFer}) (straight lines).
Here, the Hamiltonians Eq. (\ref{SSHHam}) and Eq. (\ref{DouFer}) are parametrized by
$H(k)=X(k)\sigma^1+Y(k)\sigma^2$.
The Berry phase is just the winding number of $(X,Y)$ around $X=Y=0$.
The two straight lines can be regarded as a deformation of the circle.
}
\label{f:SSHBer}
\end{center}
\end{figure}

In what follows, we compute the Berry phases in Eq. (\ref{DirBerPha}).
The reflection symmetry of the Hamiltonian (\ref{DouFer})
\begin{alignat}1
M_xH_\alpha(k)M_x^{-1}=H_\alpha(-k)
\end{alignat}
allows us to relate the wave functions at $-k$ and $k$ such that
 \begin{alignat}1
 \psi_{\alpha}(-k)=M_x\psi_{\alpha}(k)e^{i\theta_\alpha(k)}.
 \end{alignat}
This leads to $A_\alpha(-k)=-A_\alpha(k)-i\partial_k\theta_\alpha(k)$. It follows that
the Berry phase is given by
\begin{alignat}1
q_\alpha=\theta_\alpha(0)-\theta_\alpha(\infty).
\label{BerPhaDir}
\end{alignat}
The phase $\theta_\alpha(k)$ has a constraint at the reflection invariant momentum $k=0$:
\begin{alignat}1
p_\alpha e^{i\theta_\alpha(0)}=1,
\end{alignat}
where $p_\alpha=\pm1$ stands for the parity of the ground state wave function
$M_x\psi_\alpha(0)=p_\alpha\psi_\alpha(0)$, i.e., 
$p_\alpha=-{\rm sgn} (\gamma\pm1)$
for the Hamiltonian Eq. (\ref{DouFer}). Thus, we have 
\begin{alignat}1
\theta_\alpha(0) =\pm\frac{1+{\rm sgn }(\gamma\pm1)}{2}\pi,
\label{BerPhaDir0}
\end{alignat}
where the prefactor $\pm$ in the right-hand side has been introduced for later convenience.
It is possible because of  $\theta_\alpha(0)=0,\pm\pi$ mod $2\pi$.
On the other hand, the momentum space for the continuum Dirac model (\ref{DouFer}) 
is open at $k=\pm\infty$, there is no constraint on $\theta_\alpha(\infty)$. 
Thus, the Berry phase (\ref{BerPhaDir}) has ambiguity due to $\theta_\alpha(\infty)$.

However, it is noted that
from Eq.  (\ref{DouFer}), $H_0(\pm\infty)=H_\pi(\mp\infty)$ holds.
See also Fig. \ref{f:SSHBer}. 
It is thus natural
to regard the two straight lines are on single points at $k=\pm\infty$, and hence, to
choose the wave functions $\psi_0(\pm\infty)=\psi_\pi(\mp\infty)$.  
Then, the phase $\theta_0(\infty)+\theta_\pi(\infty)=0$ mod $2\pi$, and we reach
\begin{alignat}1
q_{\rm SSH}&=\theta_0(0)+\theta_\pi(0) 
\nonumber\\
&=\frac{\pi}{2}\big[{\rm sgn }(\gamma+1)-{\rm sgn }(\gamma-1) \big]
\nonumber\\
&=\left\{
\begin{array}{ll}\pi\quad &(|\gamma|<1)\\0&(|\gamma|>1)
\end{array}
\right. \mbox{mod }2\pi .
\label{DouBer}
\end{alignat}
Therefore, the lattice model has the Berry phase $q_{\rm SSH}=\pi$ 
only when $|\gamma|<1(=|\lambda|)$ \cite{Ryu:2002fk}.


\subsection{Edge states}
\label{s:edge}

Next, let us discuss the edge states of the model with boundaries described by the
Hamiltonian corresponding to Eq. (\ref{DouFer})
\begin{alignat}1
H_\alpha(x)=\mp i\sigma^2\partial_x+\sigma^1(\gamma\pm1),\quad \alpha=0,\pi.
\label{DouFerX}
\end{alignat}
As normalizable edge states at $x=0$, 
we have the following candidates 
\begin{alignat}1
\psi_{\alpha,-}(x)\propto
\begin{pmatrix}e^{\mp(\gamma\pm1)x}\\0\end{pmatrix},
\quad
\psi_{\alpha,+}(x)\propto
\begin{pmatrix}0\\e^{\pm(\gamma\pm1) x}\end{pmatrix} .
\label{SSHEdgSta}
\end{alignat}
The reflection symmetry 
\begin{alignat}1
M_xH_\alpha(x)M_x^{-1}=H_\alpha(-x),
\label{RefSymX}
\end{alignat}
ensures
\begin{alignat}1
M_x\psi_{\alpha,\mp}(x)=\psi_{\alpha,\pm}(-x).
\end{alignat}

When $q_{\rm SSH}=\pi$, the model should show the edge state located at the boundary.
Let us examine the edge state form the point of view of the Dirac fermion and its doubler
described by $H_0$ and $H_\pi$.
To this end,
let us cut the one-dimensional chain at the dashed-line in Fig. \ref{f:SSH}.
This is equivalent to imposing the conditions on the wave function of the lattice model, 
$\psi_{j=1,-}=
\psi_{ j=0,+}=0$.
According to Eq. (\ref{WavFun}),
this boundary condition is translated into 
$\psi_{0,-}(0)-\psi_{\pi,-}(0)=0$
and $\psi_{0,+}(0)+\psi_{\pi,+}(0)=0$ in the continuum limit.

Let us first consider the case $|\gamma|<1$. 
Both $H_0$ and $H_\pi$ allows
the zero energy edge states on the $x>0$ semi-infinite line, 
so we have
\begin{alignat}1
\psi_{\rm L}(x)\sim
\alpha\begin{pmatrix} e^{-(\gamma+1) x}\\0\end{pmatrix}
+e^{i\frac{\pi}{a}x}\beta \begin{pmatrix}e^{(\gamma-1) x}\\0\end{pmatrix},
\label{Lef}
\end{alignat}
where $\rm L$ means that the state is localized at the left edge $x=0$ of $x>0$. Although
this state has two independent parameters $\alpha$ and $\beta$, 
the above boundary  conditions give one constraint
$\alpha-\beta=0$, and one remaining parameter is determined by the normalization. 
Thus, only one edge state is allowed.
This state has indeed definite chirality.

On the other hand, when $1<\gamma$, we have the general solution on $x>0$,
\begin{alignat}1
\psi_{\rm L}(x)\sim
\alpha\begin{pmatrix} e^{-(\gamma+1) x}\\0\end{pmatrix}
+e^{i\frac{\pi}{a}x}\beta \begin{pmatrix}0\\e^{-(\gamma-1) x}\end{pmatrix}.
\end{alignat}
This also includes two parameters, but the boundary conditions give $\alpha=\beta=0$. 
The case $\gamma<-1$ is likewise.
Therefore, it turns out that no edge states are allowed for $|\gamma|>1$.
We here conclude that 
the edge states of the lattice model  are actually determined by the combination of those of 
a Dirac fermion and its doubler (\ref{SSHEdgSta}).

It should also be noted that in the case $|\gamma|<1$
the edge state on the other semi-infinite line $x<0$ is given by
\begin{alignat}1
\psi_{\rm R}(x)\sim
\alpha\begin{pmatrix}0\\e^{(\gamma+1) x}\end{pmatrix}+\beta e^{i\frac{\pi}{a}x}
\begin{pmatrix}0\\e^{(\gamma-1) x}\end{pmatrix}.
\label{Rig}
\end{alignat}
The two edge states $\psi_{\rm L}$ (\ref{Lef}) and $\psi_{\rm R}$ (\ref{Rig}) are related each other by the reflection symmetry (\ref{RefSymX})
\begin{alignat}1
M_x\psi_{\rm L}(x)=\psi_{\rm R}(-x).
\end{alignat}
Therefore, if the system is defined on a finite chain, e.g., $-\ell<x<\ell$, degenerate edge states
are allowed at each edge.

\subsection{Symmetry breaking perturbations}
\label{s:sym_breaking}
So far we have examined the Dirac fermion with reflection symmetry as well as chiral symmetry.
Within two-band system, one may introduce a term such as $H'=\eta \sigma^3 \sin k$ into Eq. (\ref{SSHHam})
to break chiral symmetry with keeping reflection symmetry.
Even in this model, the degenerate zero energy edge states $\psi_{\rm L,R}$ survive due to particle-hole symmetry.

\begin{figure}[htb]
\begin{center}
\includegraphics[scale=0.75]{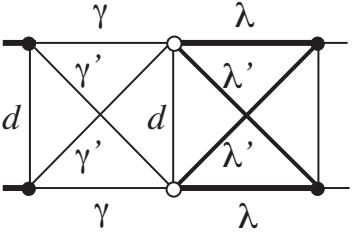}
\caption{
The 2-leg SSH ladder model which has reflection (inversion) symmetry but broken chiral symmetry.
}
\label{f:SSHlad}
\end{center}
\end{figure}

To demonstrate the model without chiral symmetry and particle-hole symmetry, 
let us consider a simple  ladder generalization of the SSH model illustrated in  Fig. \ref{f:SSHlad}.
The Hamiltonians of 2- and 3-leg ladder system are given by
\begin{alignat}1
&H_{\rm 2leg}=
\begin{pmatrix}
H_{\rm SSH}& H_{\rm SSH}'+d\1\\
H_{\rm SSH}'+d\1&H_{\rm SSH} 
\end{pmatrix},
\nonumber\\
&H_{\rm 3leg}=
\begin{pmatrix}
H_{\rm SSH}& H_{\rm SSH}'+d\1&0\\
H_{\rm SSH}'+d\1&H_{\rm SSH}&H_{\rm SSH}'+d\1 \\
0&H_{\rm SSH}'+d\1&H_{\rm SSH} 
\end{pmatrix},
\label{SSHLad}
\end{alignat}
where $H_{\rm SSH}$ and $H_{\rm SSH}'$ are given by
Eq. (\ref{SSHHam}) with intra-chain parameters $\gamma$
and $\lambda$ and  with inter-chain parameters
$\gamma'$ and $\lambda'$, respectively.
This ladder model has reflection symmetry implemented by $M_x=\sigma^1\otimes1$
but does not have chiral symmetry and particle-hole symmetry.

Suitable change of the basis, the Hamiltonian (\ref{SSHLad})
can be converted into
\begin{alignat}1
H_{\rm 2leg}=
{\rm diag} \big(&
H_{\rm SSH}+H_{\rm SSH}'+d\1,H_{\rm SSH} -H_{\rm SSH}'-d\1\big),
\nonumber\\
H_{\rm 3leg}=
{\rm diag} \big(&
H_{\rm SSH}+\sqrt{2}(H_{\rm SSH}'+d\1),H_{\rm SSH},
\nonumber\\
&H_{\rm SSH} -\sqrt{2}(H_{\rm SSH}'+d\1)\big)
\label{SSHLad2}
\end{alignat}
Since $H_{\rm SSH} \pm H_{\rm SSH}'$ is also the SSH model, 
it thus turns out that the 2-leg ladder model has edge states when 
$|\gamma\pm\gamma'|<|\lambda\pm\lambda'|$ {\it at non-zero energies $E=\pm d$.}
Interestingly, although the 3-leg ladder model has broken chiral symmetry, the edge states 
at half-filling is controlled by $H_{\rm SSH}$.
Thus, in the continuum limit of Eq. (\ref{SSHLad}) or (\ref{SSHLad2}), 
the ladder Hamiltonian shows several Dirac fermions not necessarily at zero energy.
However, these Dirac Hamiltonians should be given by Eq. (\ref{DouFer})  locally near the Fermi energy
if the system has reflection symmetry,
and the $2\times2$ Dirac fermions with chiral symmetry
are responsible for the edge states.

\subsection{A single Dirac fermion}
\label{s:single}

So far we have argued that 
the doubling of the Dirac fermion gives the correct quantized Berry phase 
of the SSH model (\ref{DouBer}) and corresponding edge states (\ref{Lef}), 
although the topological property of the edge state of each Dirac fermion (\ref{SSHEdgSta})
with a sharp boundary is not clear. 
To reveal this,
it may be convenient to introduce smooth boundary 
%
connecting different mass parameters at $x=\pm\infty$.
In this case, edge states can be regarded  as domain wall states to which topological invariant can be assigned. 

To demonstrate this, we focus our attention to the topological property of a single Dirac fermion
with the mass parameter $\phi$, i.e.,
\begin{alignat}1 
H=-i\Gamma^1\partial_x+\Gamma^2\phi(x) ,
\label{SinDirHam}
\end{alignat}
where we choose
$\Gamma^1=\sigma^2$, $\Gamma^2=\sigma^1$, and $\Gamma_5=-\sigma^3$ for
$H_0$,  and
$\Gamma^1=-\sigma^2$, $\Gamma^2=-\sigma^1$, and $\Gamma_5=-\sigma^3$ for
$H_\pi$ in Eq. (\ref{DouFerX}) to ensure 
\begin{alignat}1
\tr \Gamma_5\Gamma^1\Gamma^2=2i.
\label{ChoGam}
\end{alignat}
This is the famous Jackiw-Rebbi model \cite{Jackiw:1976fk}.
We here assume that $\phi$ is a smooth function of $x$, and becomes constant 
at $x\rightarrow\pm\infty$, $\phi(\pm\infty)=$const.
When $\phi(-\infty)<0<\phi(\infty)$, for example, we have the normalizable edge state,
\begin{alignat}1
\psi_{-}\propto\begin{pmatrix}e^{-\int^x\phi(x')dx'}\\0\end{pmatrix},
\end{alignat}
with chirality $\Gamma_5=-1$, and hence $n_-=1$, $n_+=0$ and ${\rm ind}~H=-1$
in Eq. (\ref{IndTheOri}).
This state is a smooth extension of the edge state $\phi_{0,-}(x)$ in Eq. (\ref{SSHEdgSta}) towards $x<0$.
Likewise, when $\phi(\infty)<0<\phi(-\infty)$, we have 
$n_-=0$, $n_+=1$ and ${\rm ind}~H=+1$,
and in other cases where $\phi(\pm\infty)$ have the same signs, we have no edge states, and 
hence ${\rm ind}~H=0$.

Because of chiral symmetry, the edge states or domain wall states are at zero energy. In this case,
the index theorem clarifies the direct equivalence between the analytical index associated with the zero energy
domain wall  states and topological index associated with the Berry phase.
In this subsection, we briefly investigate the index theorem Eq. (\ref{IndTheOri}) 
for the present Dirac Hamiltonian (\ref{SinDirHam}).
In one dimension, the right-hand side of Eq. (\ref{IndThe}) is modified into
\begin{alignat}1
{\rm ind }~H&=
-\frac{1}{2}\int \partial_x\langle j_5^1\rangle dx
\nonumber\\
&=
-\frac{1}{2}\left[\langle j_5^1\rangle(x=\infty)-\langle j_5^1\rangle(x=-\infty)\right].
\label{CurInt}
\end{alignat}
The current $\langle j_5^1\rangle(x)$ 
can be calculated in a similar way to Eqs. (\ref{AxiCur}), (\ref{PlaWavHam}), and (\ref{DerExp}), 
\begin{alignat}1
\langle j_5^1(x)\rangle&=\int\frac{dk}{2\pi}{\rm tr }\,(\Gamma^3)\Gamma^1
\frac{-\Gamma^1(\partial_x+i k)-i\Gamma^{2}\phi(x)}{-(\partial_x+ik)^2+\phi^2-\sigma^3\partial_x\phi}
\nonumber\\
&=\int\frac{dk}{2\pi}
\frac{2\phi(x)}{k^2+\phi^2(x)}
={\rm sgn} ~\phi(x) ,
\label{CurExp}
\end{alignat}
We finally obtain the topological index
\begin{alignat}1
{\rm ind }~H
&=-\frac{1}{2}\left[{\rm sgn}\, \phi(\infty)-{\rm sgn}\, \phi(-\infty)\right],
\label{SSHInd}
\end{alignat}
which is exactly the same as the analytical index mentioned above.
Note that this result is nothing to do with specific choices of the $\Gamma$-matrices in Eq. (\ref{SinDirHam}):
Only the anti-commutation relation between $\Gamma$-matrices 
and the normalization in Eq. (\ref{ChoGam}) are responsible for Eqs. 
(\ref{CurExp}) and (\ref{SSHInd}).
Since the index of $H$ may be given by the topological numbers, we assume
\begin{alignat}1
{\rm ind }~H=-[q(\infty)-q(-\infty)]/\pi .
\label{IndBer}
\end{alignat}
Then, we have 
$
q(\pm\infty)=\frac{\pi}{2}{\rm sgn}\,\phi(\pm\infty)
$
apart from a constant.
Thus, for the fermion with a constant mass $m$, $H=-i\Gamma^1\partial_x+\Gamma^2 m$, 
it is reasonable to assign the topological number $q=\frac{\pi}{2}{\rm sgn}\,m$.
This is nothing but the Berry phase with a special choice of gauge.
With the definition of the $\Gamma$-matrices in Eq. (\ref{SinDirHam}) for $H_\alpha$, 
the masses  are $1+\gamma$ and
$1-\gamma$ for $\alpha=0$ and $\alpha=\pi$, respectively.
This leads to $q_0=\frac{\pi}{2}{\rm sgn}\,(1+\gamma)$ and $q_\pi=\frac{\pi}{2}{\rm sgn}\,(1-\gamma)$,
which indeed reproduces Eq. (\ref{DouBer}).
Uncertain constant terms vanish if the Dirac fermions and its doublers are combined.



\end{document}